\documentclass[a4paper,11pt]{amsart}
\usepackage{amssymb}
\begin{document}

\hyphenation{gra-vi-ta-tio-nal re-la-ti-vi-ty Gaus-sian
re-fe-ren-ce re-la-ti-ve gra-vi-ta-tion Schwarz-schild
ac-cor-dingly gra-vi-ta-tio-nal-ly re-la-ti-vi-stic pro-du-cing
de-ri-va-ti-ve ge-ne-ral ex-pli-citly des-cri-bed ma-the-ma-ti-cal
de-si-gnan-do-si coe-ren-za pro-blem gra-vi-ta-ting geo-de-sic
per-ga-mon cos-mo-lo-gi-cal gra-vity cor-res-pon-ding
de-fi-ni-tion phy-si-ka-li-schen ma-the-ma-ti-sches ge-ra-de
Sze-keres con-si-de-red tra-vel-ling ma-ni-fold re-fe-ren-ces
geo-me-tri-cal in-su-pe-rable sup-po-sedly at-tri-bu-table}

\title[On some relativistic singular surfaces]
{{\bf On some relativistic singular surfaces}}

\author[Angelo Loinger]{Angelo Loinger}
\address{A.L. -- Dipartimento di Fisica, Universit\`a di Milano, Via
Celoria, 16 - 20133 Milano (Italy)}
\author[Tiziana Marsico]{Tiziana Marsico}
\address{T.M. -- Liceo Classico ``G. Berchet'', Via della Commenda, 26 - 20122 Milano (Italy)}
\email{angelo.loinger@mi.infn.it} \email{martiz64@libero.it}

\vskip0.50cm

\begin{abstract}
In the current relativistic literature there are misleading
considerations about some singular surfaces. An accurate geometric
analysis allows to settle the question. No physical meaning is
attributable to the spatial regions surrounded by the above
surfaces. \\A recent observational paper by A. Fabian \emph{et
al.} is discussed in sect. \textbf{4}.
\end{abstract}

\maketitle


\normalsize

\vskip1.20cm \noindent \textbf{1.} -- Levi-Civita has given a
beautiful, \emph{geometrically detailed} treatment of the
Schwarzschild manifold created by a gravitating point-mass $m$
\cite{1}. First of all, following a Palatini's method \cite{2},
Levi-Civita makes a precise illustration of the geometrical
concepts that suggest how to choose a convenient \emph{system of
coordinates} in a three-dimensional metric space $V_{3}$ endowed
with a spherical symmetry around one of its points $O$. He
institutes a one-to-one correspondence between $V_{3}$ and an
auxiliary three-dimensional Euclidean space $V_{3}'$ (a
\emph{Bildraum}, according to Weyl's terminology), having a
spherical symmetry around a point $O\,'$. Then, he proves that at
any \emph{ray} $j\,'$ starting from $O\,'$ there corresponds in
$V_{3}$  a \emph{geodesic line} $j$ starting from $O$; and that at
any \emph{spherical surface} $\Sigma\,'$ of $V_{3}'$, with centre
at $O\,'$, there corresponds in $V_{3}$ a \emph{geodesic sphere}
$\Sigma$ with centre at $O$. Let $\textrm{d}\sigma\,'$ be an
elementary interval on $\Sigma\,'$ drawn from one of its points
$Q\,'$, and $\textrm{d}\sigma$ the homologous interval drawn from
the corresponding point $Q$ on $\Sigma$. If we refer the
\emph{Bildraum} $V_{3}'$ to polar coordinates $r, \vartheta,
\varphi$, we have

\begin{equation} \label{eq:one}
\textrm{d}\sigma\,'^{2} = r^{2} (\textrm{d}\vartheta^{2} +
\sin^{2}\vartheta \, \textrm{d}\varphi^{2}) \quad; \quad
r=O\,'Q\,' \quad;
\end{equation}

owing to the correspondence between $V_{3}'$ and $V_{3}$, the
coordinates $\vartheta$ and $\varphi$ can be also considered as
\emph{curvilinear coordinates on} $\Sigma$. It is easy to prove
that the correspondence between $\Sigma$ and $\Sigma\,'$ is a
\emph{conformal} one, \emph{i.e.} that there exist a function
$H(r)$ such that

\begin{equation} \label{eq:two}
\textrm{d}\sigma^{2} = H^{2}(r) \, \textrm{d}\sigma\,'^{2} \quad ;
\end{equation}

if $R(r) := r H(r)$, the square of the line element
$\textrm{d}\sigma$ of $\Sigma$ can be written

\begin{equation} \label{eq:three}
\textrm{d}\sigma^{2}  = R^{2}(r) \, (\textrm{d}\vartheta^{2} +
\sin^{2}\vartheta \, \textrm{d}\varphi^{2}) \quad.
\end{equation}

In a second Euclidean \emph{Bildraum}, eq. (\ref{eq:three}) gives
the elementary interval on a sphere of radius $R(r)$, of which
$K:=1/R^{2}$ is the Gaussian curvature, and $S:=4\pi R^{2}$ the
superficial area: two \emph{invariant} concepts. Consequently, in
the metric space $V_{3}$ the \emph{coordinate} $R$ has the
following meaning: at any point $P$, $1/R^{2}$ represents the
Gaussian curvature of the geodesic sphere $\Sigma$ with centre
$O$, passing through $P$. Further, since all geodesic lines
starting from $O$ cut orthogonally $\Sigma$, the interval
$\textrm{d}l$ of $V_{3}$ is given by

\begin{equation} \label{eq:four}
\textrm{d}l^{2} = \textrm{d}g^{2} + \textrm{d}\sigma^{2} \quad,
\end{equation}

where $\textrm{d}g$ is the elementary arc of one of the geodesics
-- arc which depends only on $R$:

\begin{equation} \label{eq:five}
\textrm{d}g = A(R) \, \textrm{d}R \quad;
\end{equation}

from which:

\begin{equation} \label{eq:six}
\textrm{d}l^{2} = A^{2}(R) \, \textrm{d}R^{2} +  R^{2} \,
(\textrm{d}\vartheta^{2} + \sin^{2}\vartheta \,
\textrm{d}\varphi^{2})\quad.
\end{equation}

This result holds under ``the obvious condition that the
coefficients of $\textrm{d}l^{2}$ are regular in the region round
every point, except possibly the point $O$.''

\par If $\textrm{d}s^{2} = U(R)\,\textrm{d}t^{2}-\textrm{d}l^{2}$ is
the interval of the Schwarzschild manifold, Einstein equations
$R_{jk}=0$, $(j,k=0,1,2,3)$, tell us that $(c=G=1)$:

\begin{equation} \label{eq:seven}
\textrm{d}s^{2} =  \left(1-\frac{2m}{R(r)} \right) \,
\textrm{d}t^{2} -
\frac{\left[\textrm{d}R(r)\right]^{2}}{\left(1-\frac{2m}{R(r)}\right)}
-  R^{2}(r) \, (\textrm{d}\vartheta^{2} + \sin^{2}\vartheta \,
\textrm{d}\varphi^{2}) \quad;
\end{equation}

if we put $R(r)\equiv r$, we obtain the \emph{standard}
(Hilbert-Droste-Weyl) form of $\textrm{d}s^{2}$. For $R(r)\equiv
[r^{3}+(2m)^{3}]^{1/3}$, we have the \emph{original }Schwarzschild
form (1916). For $R(r)\equiv r+2m$, the Brillouin form (1923).
\emph{Etc.} -- By virtue of eq. (\ref{eq:six}), $1-2m/R(r)$ cannot
be negative. Remark that Schwarzschild's and Brillouin's metrics
satisfy the above obvious condition; the standard metric holds
only for $r>2m$.

\vskip1.20cm \noindent \textbf{2.} -- We have given a faithful
summary of the first part of Levi-Civita's argumentation, because
it evidences the geometric meaning of the radial coordinate
$R(r)$.

\par The \emph{distance} $D$ of a generic point $R(r)=\varrho$
from $R(r)=2m$ is obviously given by

\begin{eqnarray} \label{eq:eight}
D & = & \int_{2m}^{\varrho}
\left[\frac{R(r)}{R(r)-2m}\right]^{1/2} \textrm{d}R(r) =
\nonumber\\
& & {} 
=  \left\{ R(r) \, [R(r)-2m]\right\} ^{1/2} + m \,
 \ln \frac{R(r)-m+\left\{R(r) \, [R(r)-2m]\right\}^{1/2}}{m}\quad;
\end{eqnarray}

and \emph{since} $R(r)/[R(r)-2m]=A^{2}\,[R(r)]$, \emph{we see
that} $R(r)>2m$.

\par Obviously, the ``soft'' singularity of interval
(\ref{eq:seven}) at $R(r)=2m$ does not have a radial
\emph{distance} $d=2m$ from $R(r)=0$. The popular formula

\begin{equation} \label{eq:nine}
(S/4\pi)^{1/2} \approx 3 \, \textrm{km} \, \,
\frac{\textrm{point-mass }m}{\textrm{solar mass }}
\end{equation}

gives only a formal value of the radial \emph{coordinate}. But the
essential fact is that for $R(r)\leq 2m$ metric (\ref{eq:seven})
loses any validity; consequently, \emph{space region} $R(r)\leq
2m$ \emph{does not admit of any reasonable interpretation}. In
particular, formula (\ref{eq:nine}) is problematic.

\par Some authors have asserted that when $R(r)<2m$ the spatial
coordinate $R(r)$ could acquire a time character and the temporal
coordinate $t$ could acquire a space character.  But this change
of steed would give a $\textrm{d}s^{2}$ which has \emph{no}
relation with our \emph{static} problem -- as it was irrefutably
demonstrated many years ago by Brillouin \cite{3}.

\vskip1.20cm \noindent \textbf{2bis.} -- The second fundamental
memoir of GR by Schwarzschild (1916) \cite{4} ends with this
remark: a sphere F of an incompressible fluid, having a
\emph{given} gravitating mass $m$, has a minimal value of its
\emph{physical radius} which is equal to $(9/8)(2m)$: a result
with a real meaning, because it was obtained through the
concordance at the surface of F between external and internal
$\textrm{d}s^{2}$, \emph{in absence of any whatever singularity}.

\vskip1.20cm \noindent \textbf{3.} -- Of course, for Kerr's metric
\cite{5} and for the metric of a gravitating mass-point with an
electric charge \cite{6} we can make considerations very similar
to those of sect. \textbf{2}.

\vskip1.20cm \noindent \textbf{4.} -- A test-particle or a
light-ray moving along a radial geodesic of Schwarzschild manifold
cannot overcome the ``barrier'' at $R(r)=2m$, owing to Hilbert
\emph{repulsive} effect \cite{7}. A quite analogous result holds
for Kerr's metric \cite{5} and for the metric of a charged
mass-point \cite{6}.

\par Accordingly, \emph{the involved singularities cannot swallow
anything}. However, in the recent literature we find many
assertions that the authors have detected various voracious eaters
of matter and light, that they identify with BH's -- see
\emph{e.g.} the paper by Fabian \emph{et al.} \cite{8}, which
would concern a supermassive BH. This team of astrophysicists has
observed a broad iron $K$ and $L$ line emission in the Narrow-Line
Seyfert 1 Galaxy 1H0707-495, using the XMM-Newton satellite. The
observational data tell that this Galaxy is  spinning very rapidly
and is swallowing the equivalent of two Earths per hour. Fabian
\emph{et al.} \cite{8} think that at the galactic heart there is a
supermassive BH, whose mass is estimated at about 3 to 5 million
solar masses. The researchers would be tracking matter to within
twice the ``radius'' of the hypothesized BH -- \emph{evidently,
they utilize formula} \cite{9}. According to Fabian, this BH is a
messy eater, since ``accretion is a very messy process because of
 the magnetic fields that are involved.''

 \par Now, Wolfgang Kundt, a distinguished astrophysicist who does
 not object to the theoretical belief in the existence of BH's,
 has repeatedly emphasized that there are very realistic
 explanations of the observational data supposedly regarding
 various kinds of BH's, in particular the supermassive ones -- see
 \emph{e.g.} his concise \emph{Astrophysics} \cite{9}.

 \par He remarks that the active galactic nuclei (AGN), that are
 commonly thought to harbour a supermassive BH,``may, instead, simply be
 the high-density nuclear-burning centers of galactic disks.''
 Indeed ``nuclear burning (of H to Fe) is almost as efficient a lamp as black-hole
 accretion, yielding $\lesssim 1\%$ of the rest-energy'', as it is suggested by nine
 reasons (pp.101, 102 of \cite{9})! And further: ``The QSO phenomenon may rather
 be a straightforward consequence of the spiralling-in of matter through \emph{galactic disks} whose
 density approaches stellar values near their centers, giving rise to almost
 relativistic Kepler velocities on the innermost \emph{Solar-System scales},
 to strong magnetization and magnetic reconnections, and to nuclear burning
 in the disk's midplane. Magnetic reconnections create the pair plasma which
 is responsible for the jet phenomenon and for the hard, non thermal spectra,
 whereas nuclear power is used to re-eject matter at higher than SN velocities through the BLR.
 Active galactic nuclei may owe their extreme properties to those of their central disks.''

\vskip1.20cm \noindent \textbf{5.} -- \emph{Two final remarks}.
--\emph{i}) The original Schwarzschild's form of the
$\textrm{d}s^{2}$, which corresponds to $R(r)\equiv
[r^{3}+(2m)^{3}]^{1/3}$, and Brillouin's form, which corresponds
to $R(r)\equiv r+2m$, are the \emph{unique} maximally extended
(and geodesically complete) metric forms of Schwarzschild manifold
that do not impair, or partially suppress, the gravitational field
of the point-mass. --\emph{ii}) It is commonly believed (without a
\emph{rigorous} proof) that since $p/c^{2}$ gives a contribution
to the mass density of any celestial body, even infinite pressure
gradients cannot support a star of a sufficiently great mass
against its self-attraction, with the formation of a BH as a final
outcome. Now, this opinion is not right, as McVittie \cite{10} and
the present authors \cite{11} have demonstrated with precise
computations; a result which corroborates a famous conjecture by
Eddington.


\end{document}